          [2004/06/22 1.2 (KOF); 2001/04/25 1.1 (PWD)]

\documentclass[letter,twocolumn]{Gaia2004} 
\usepackage{times}      
\usepackage{epsfig}     
\usepackage{natbib}     
\title{Self-consistent distance determinations for Lutz-Kelker-limited samples}

\author[1,2]{Jes\'us Ma{\'{\i}}z Apell\'aniz}
\affil[1]{STScI, Baltimore, Maryland, U.S.A.}
\affil[2]{Space Telescope Division, ESA, ESTEC, Netherlands}

\bibpunct{(}{)}{;}{a}{}{,}  

\begin{document}


\maketitle

\begin{abstract}
  
	We present a method designed to correct for Lutz-Kelker effects in distance-limited samples. The
method allows for the calculation of distances to individual objects and, at the same time, provides a fit to
a parameterized, self-consistent spatial distribution of the population. An example using Hipparcos data is
presented and the relevance to Gaia is also discussed.

\end{abstract}

\section{Why?}

	The measurement of distances from trigonometric parallaxes is complicated by the existence of the
Lutz-Kelker bias (Lutz \& Kelker 1973), which generates selection effects when analyzing a given stellar
population and also requires the use of correction factors for the distance to a given star derived from its
parallax (Smith 2003). Standard Lutz-Kelker corrections become significant for observed parallaxes with 
$\varepsilon_\pi \equiv \sigma_\pi/\pi_{\rm o} \ge$ 0.05 and diverge for $\varepsilon_\pi \ge$ 0.175
(Fig.~1, left panel). Ignoring them can introduce gross errors in any derived quantity. For stars observed
with Gaia, unobscured G dwarfs are expected to have $\varepsilon_\pi =$ 0.05 at $\approx$ 2 kpc; that value 
is achieved for unobscured M dwarfs at distances of $\sim$ 500 pc. It is clear that Lutz-Kelker corrections
will have to be applied to Gaia parallaxes for those objects, among others.

\begin{figure*}[H]
\begin{center}
\leavevmode
\includegraphics*[width=0.47\linewidth]{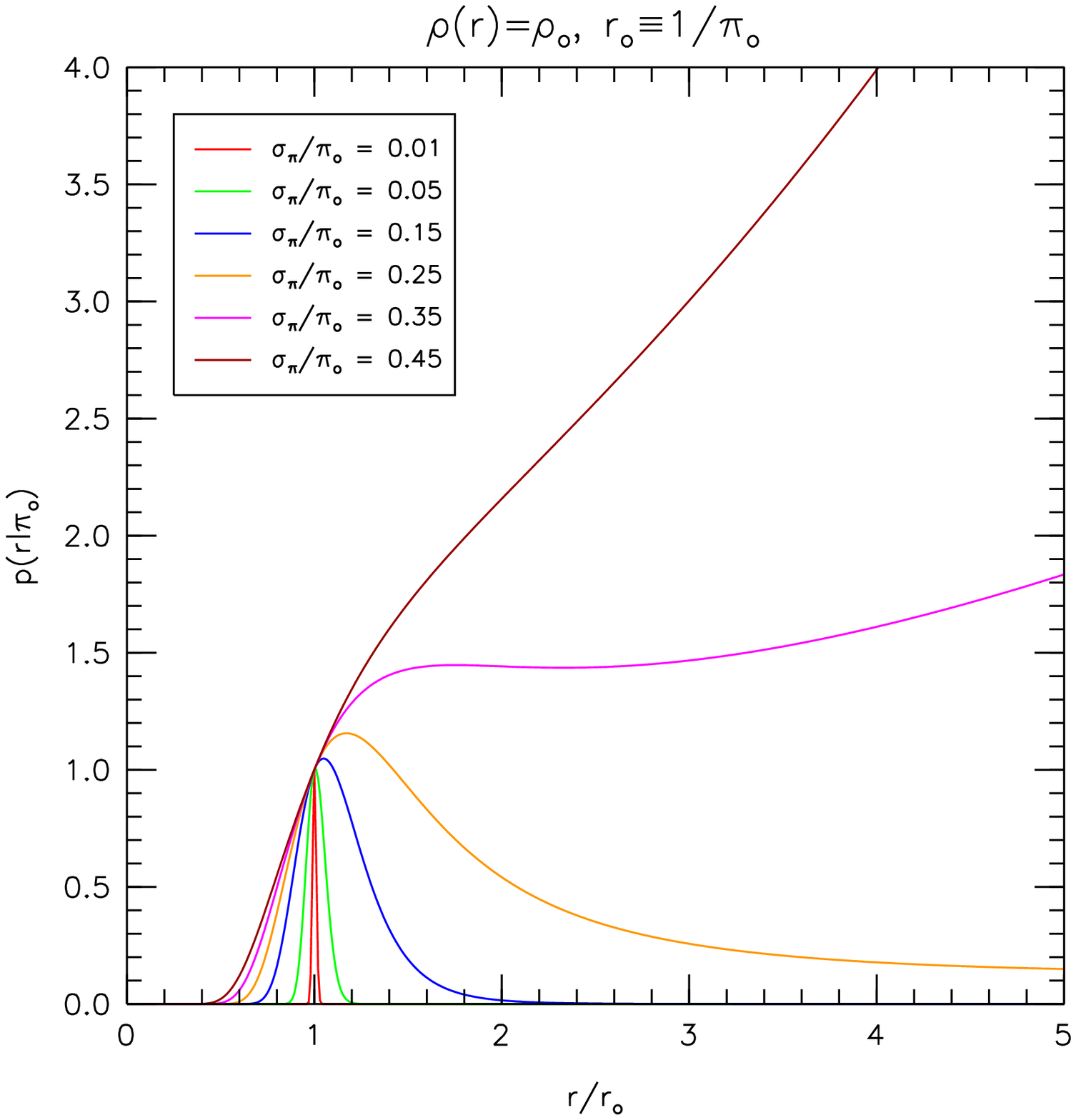} \ 
\includegraphics*[width=0.47\linewidth]{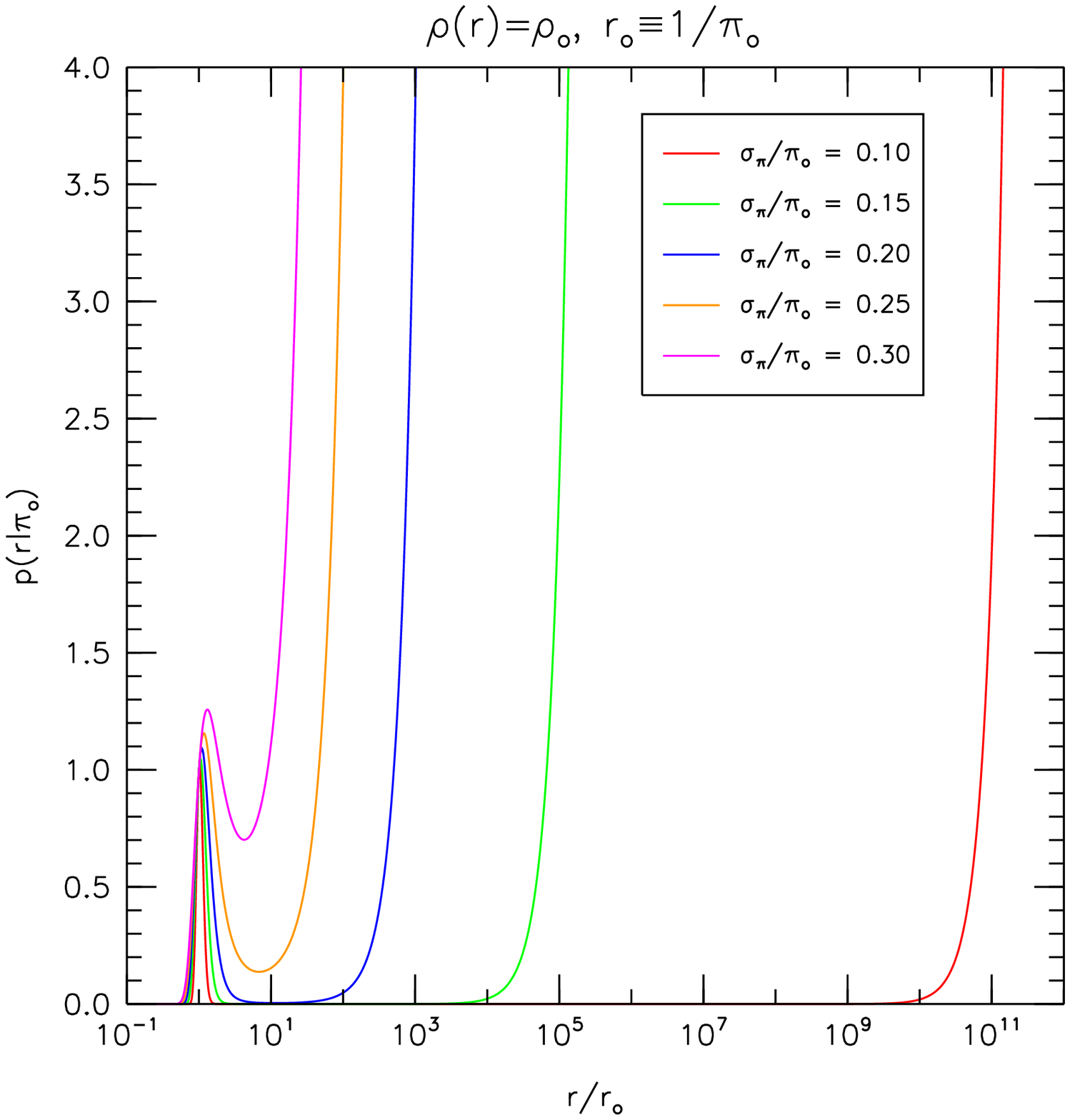}
\end{center}
\caption{Real distance probability distributions for Gaussian uncertainties
and an underlying constant spatial distribution assuming different values of $\varepsilon_\pi$. 
The left panel has a linear horizontal scale and the right one a logarithmic one.}
\end{figure*}

	The problem is even more serious than that: if we assume a constant underlying constant spatial 
distribution $\rho(r)$ and a Gaussian distribution for the parallax uncertainty (i.e. the standard
assumptions for Lutz-Kelker corrections), the real distance $r$ probability distributions for individual
stars, $p_i(r)$, will always be
ill-behaved for $r\rightarrow\infty$, even for small values of $\varepsilon_\pi$ (Fig.~1, right panel). This 
characteristic precludes a precise statistical analysis of parallaxes unless cutoffs are specified for the 
Gaussian distribution.

	The situation described in the previous paragraph is actually not a realistic representation of the
Galactic stellar populations that will be sampled by Gaia. $\rho(r)$ is not expected to be constant; on the
contrary, in most cases it is expected to drop significantly beyond a relatively short distance. This
alleviates or eliminates altogether the ill behavior of $p_i(r)$ for $r\rightarrow\infty$ but requires the
use of new techniques for its precise calculation. That is the purpose of this work.

\section{How?}

	The following procedure can be used for a simple conical or biconical
volume centered on the position of the Sun. It can also be adapted with some
modifications to other simple geometries involving cone or cylinder sections
(see Fig. 2). 

\begin{figure}[H]
\begin{center}
\leavevmode
\includegraphics*[width=\linewidth]{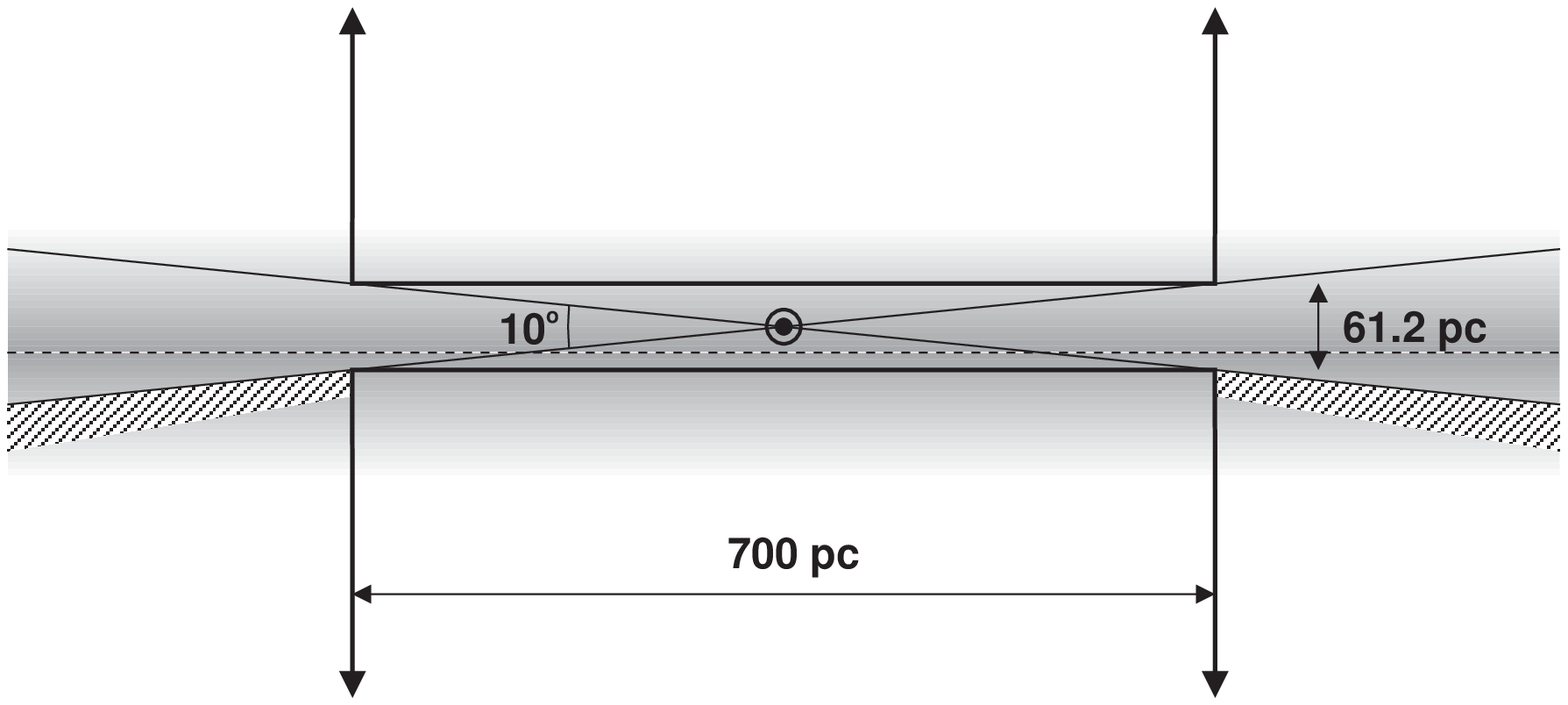}
\end{center}
\caption{The double semi-infinite cylinder used to study the 
vertical distribution of O-B5 stars is represented by the two regions 
contained inside the thick lines. The cylinder has a radius of 350 pc and the
gap is defined by excluding the stars within 5$^{\rm o}$ of the Galactic
equator. The shaded area 
represents the Galactic disk with the dashed line being the Galactic plane and 
the Sun symbol marking the position of our star. The hatched regions mark the 
area excluded to estimate the bias induced by extinction.}
\end{figure}

We start by selecting a distance-limited (not magnitude-limited) population along a given direction. It is
possible (and in most cases expected) for a good fraction of the sample to have large values of
$\varepsilon_\pi$. Negative values of $\pi_{\rm o}$ for part of the sample are also possible and, again, 
expected if the sample is severely Lutz-Kelker-limited (for example, if we are dealing with Gaia data for 
dim, low-mass stars).

We then assume that $\rho(r)$ is not constant for that population and provide a description in terms of a 
reasonable functional form that goes to zero as $r\rightarrow\infty$. For example, one can assume a Galactic
disk with constant surface density and a Gaussian profile in the vertical direction and derive the expected 
$\rho(r)$ along a given direction. More complex distributions with multiple components can also be used.

For the next step, an educated guess for the free parameters of the functional form is provided. This can be
deduced from pre-existing data. Then, the probability distribution $p_i(r)$ for each star $i$ given 
$\pi_{{\rm o},i}$, and $\sigma_{\pi,i}$ as:

\begin{equation}
p_i(r) = A\, r^2\,e^{-\frac{1}{2}\left(\frac{1-r\pi_{{\rm o},i}}
{r\sigma_{\pi,i}}\right)^2} \rho(r),
\end{equation}

where $A$ is a normalization factor.

Then we sum over all the stars in that particular direction to obtain the total probability distribution 
$p(r)$ and, if different directions are sampled, a 3-D $p(x,y,z)$ is derived. The total probability
distribution is then used to derive a $\rho(r)$ by $\chi^2$ fitting of the free parameters in the chosen
functional form.

The procedure is then iterated until convergence. At the end, the residuals of the fit can be analyzed to
check whether the selected functional form is capable of producing a reasonable representation of the data. 
If that is not the case, a new functional form can be selected and the process is repeated.

The expected quality and magnitude completeness of GAIA data will allow
the spatial distribution of different stellar populations to be analyzed with this 
method.

\section{An example}

	We have used a variation of this method with a somewhat more complex geometry
(Fig. 2) to determine the vertical structure of the spatial distribution of early-type 
stars in the solar neighborbood from Hipparcos parallaxes. The full analysis is available in 
Ma\'{\i}z-Apell\'aniz (2001a). Here we summarize the most important results:

The Sun is located above the plane of the Galaxy at a distance of
$z_\odot =$ 24.7 $\pm$ 1.7 (random) $\pm$ 0.4 (systematic) pc. This value is consistent with most of the
ones obtained by other authors using similar or different populations (Chen et al. 2001).

The scale height of the disk early-type stellar poipulation assuming a self-gravitating, isothermal,
single-mass disk is $h_s =$ 34.2 $\pm$ 0.8 (random) $\pm$ 2.5 (systematic) pc. The data does not have a good
enough precision to differentiate between that model and one in which the gravitational field is provided by
a constant-density background population, since both functional forms yield similar good values in the 
$\chi^2$ fit. The value for $h_s$ is slightly lower that the one obtained by other authors: the likely reason
for this effect is the absence of a thick disk/halo component in previous works (see below).

\begin{figure}[h]
\begin{center}
\leavevmode
\includegraphics*[width=\linewidth]{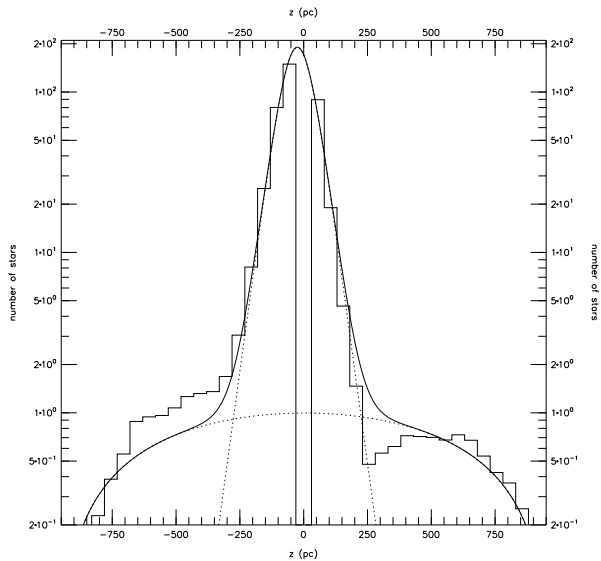}
\end{center}
\caption{Observed (histogram) and model fit frequency distributions as a
function of the vertical coordinate for O-B5 stars using a self-gravitating
isothermal disk + parabolic halo distribution. The dotted lines represent each
individual component and the continuous one the sum of the two. The displayed
bin size is uniform and equal to 50 pc, but is not used for the fit.
The region immediately surrounding $z=$ 0 is not considered for the fit, as it
corresponds to the space between the two semi-infinite cylinders in Fig.~2. Note that the 
observed distribution shows ``fractional stars'' due to the procedure used to 
derive the histogram.}
\end{figure}

The local disk surface density for O-B5 stars is
$\sigma_{\star} =$ (1.62 $\pm$ 0.04 (random) $\pm$ 0.14 (systematic))$\cdot$ 10$^{3}$ stars kpc$^{-2}$. This
is in reasonable agreement with previous studies.

A halo/thick disk component is clearly detected at large distances from the plane (Fig. 3). The dataset is 
not good enough to provide detailed information on that population except to detect its presence, infer that
it contributes with at least 5\% of the total O-B5 stellar population in our region of the Galaxy,
and deduce that it may extend beyond a distance of 500 pc from the Galactic plane. The origin of such a
component is still a matter of debate: most of those stars appear to be runaways (Rolleston et al. 1999,
Hoogerwerf et al. 2000) but some could have been formed in situ (Conlon et al. 1992).

The volume within $\sim$ 100 pc of the Sun is deficient in early-type stars. Beyond there, 
we start to find OB associations, such as Scorpius-Centaurus, that compensate the local deficiency 
(see also de Zeeuw et al. 1999, Ma\'{\i}z-Apell\'aniz 2001b).

We have used these results to include distance information derived from Hipparcos parallaxes for
some of the stars in our Galactic O Star Catalog (Ma\'{\i}z-Apell\'aniz et al. 2004). Two examples are shown 
in Fig.~4. We would like to point out that the discrepancies between trigonometric and spectroscopic
parallaxes for early-type stars claimed by Sk\'orzy\'nsky et al. (2003) and Patriarchi et al. 
(2003) are due to an incorrect treatment of Lutz-Kelker corrections; no  
discrepancies are found for the distances derived from our method.

A color version of this poster is available at {\tt http://www.stsci.edu/\~{}jmaiz}.

\begin{figure*}[h]
\begin{center}
\leavevmode
\includegraphics*[width=0.47\linewidth]{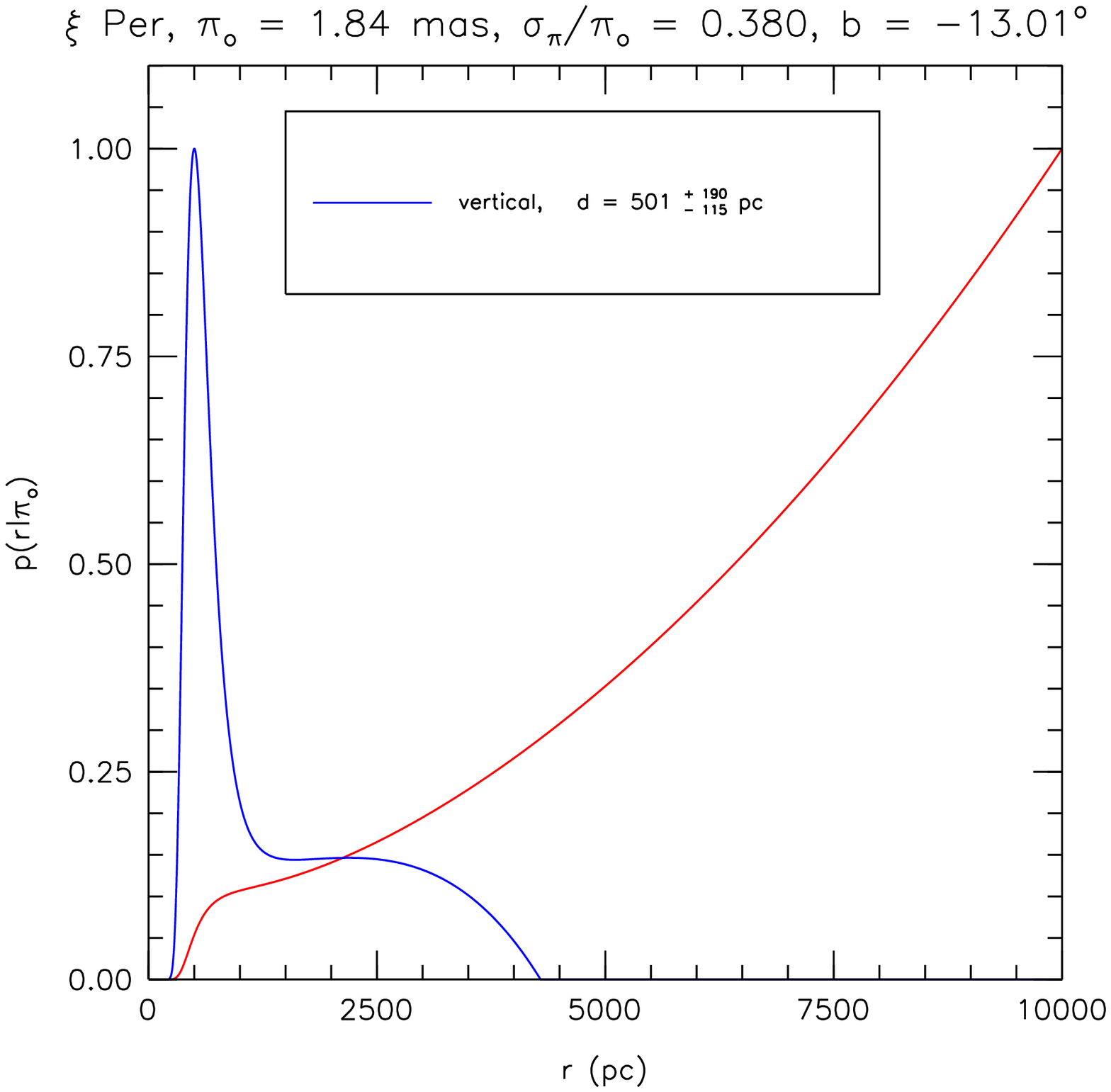} \ 
\includegraphics*[width=0.47\linewidth]{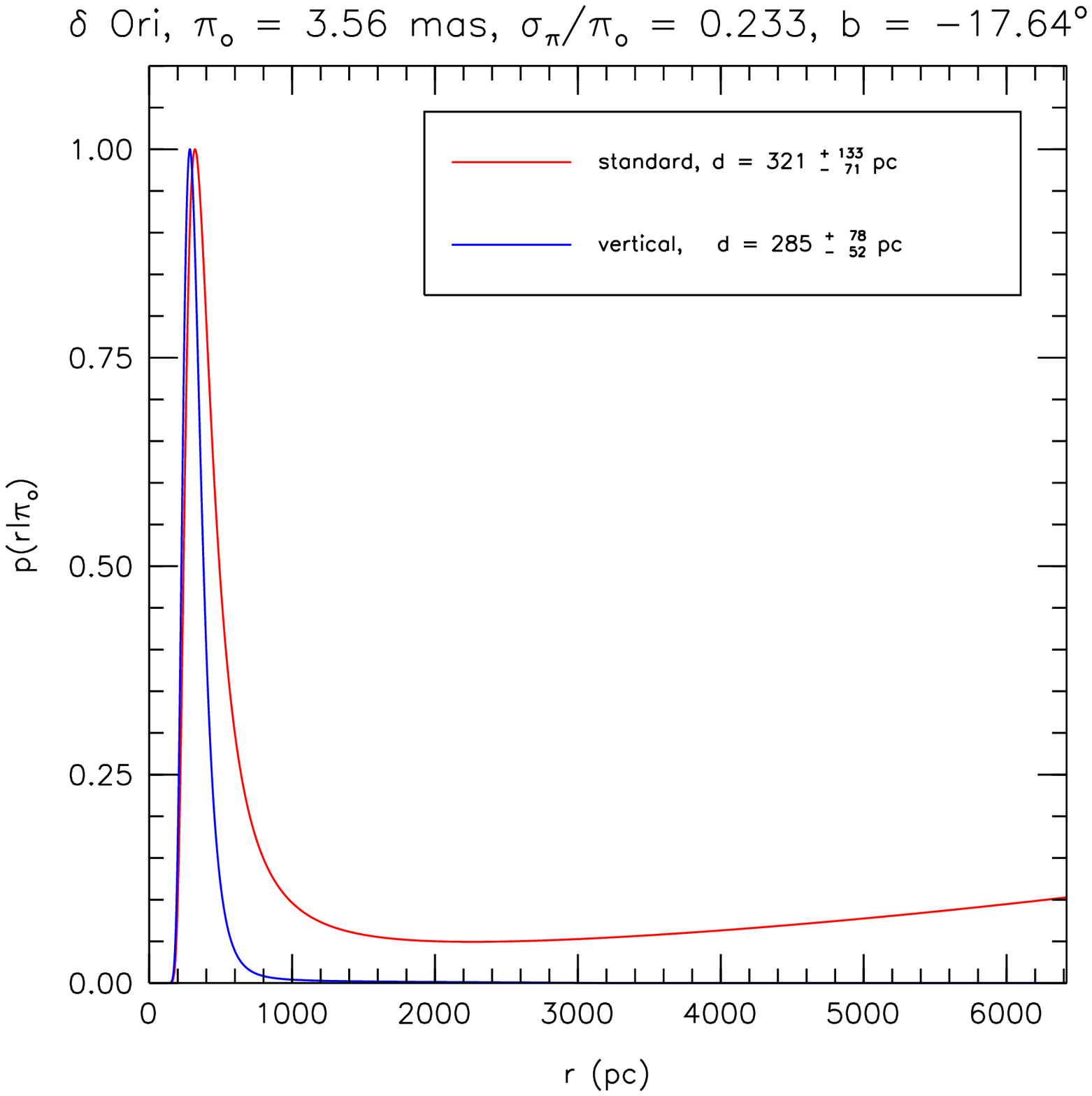}
\end{center}
\caption{Real distance probability distributions for two Galactic O stars
derived from Hipparcos parallaxes. The red curve assumes an underlying constant spatial distribution
while the the blue curve uses the Galactic vertical structure model of Ma\'{\i}z-Apell\'aniz (2001a).}
\end{figure*}




\begin{thebibliography}{}
\bibitem
[\protect\astroncite
{Chen et~al.}{2001}]{Chenetal01}
Chen, B. et al., 2001, ApJ 553, 184-197

\bibitem
[\protect\astroncite
{Conlon et~al.}{1992}]{Conletal92}
Conlon, E. S. et al., 1992, ApJ 400, 273-279

\bibitem
[\protect\astroncite
{de Zeeuw et~al.}{1999}]{deZeetal99}
de Zeeuw, P. T. et al., 1999, AJ 117, 354-399

\bibitem
[\protect\astroncite
{Hoogerwerf et~al.}{2000}]{Hoogetal00}
Hoogerwerf, R. et al., 2000, ApJL 544, 133-136

\bibitem
[\protect\astroncite
{Lutz \& Kelker}{1973}]{LutzKelk73}
Lutz, T. E. \& Kelker, D. H., 1973, PASP 85, 573-578

\bibitem
[\protect\astroncite
{Ma\'{\i}z-Apell\'aniz}{2001a}]{Maiz01a}
 Ma\'{\i}z-Apell\'aniz, J., 2001a, AJ 121, 2737-2742

\bibitem
[\protect\astroncite
{Ma\'{\i}z-Apell\'aniz}{2001b}]{Maiz01a}
 Ma\'{\i}z-Apell\'aniz, J., 2001b, ApJL 560, 83-86

\bibitem
[\protect\astroncite
{Ma\'{\i}z-Apell\'aniz et~al.}{2004}]{Maizeatl04}
 Ma\'{\i}z-Apell\'aniz, J., Walborn, N. R., Galu\'e, H. \'A., \& Wei, L. H. 2004, ApJS 151, 103-148, 
{\tt http://www.stsci.edu/\~{}jmaiz/GOSmain.html}

\bibitem
[\protect\astroncite
{Patriarchi et~al.}{2003}]{Patretal03}
Patriarchi, P. et al., 2003, A\&A 410, 905-909

\bibitem
[\protect\astroncite
{Rolleston et~al.}{1999}]{Rolletal99}
Rolleston, W. R. J. et al., 1999, A\&A 347, 69-76

\bibitem
[\protect\astroncite
{Sk\'orzy\'nskyi et~al.}{2003}]{Skoretal03}
Sk\'orzy\'nsky, W. et al., 2003, A\&A 408, 297-304

\bibitem
[\protect\astroncite
{Smith}{2003}]{Smit03}
Smith, H., 2003, MNRAS 338, 891-902

\end{thebibliography}
\end{document}